%% file: main.tex
\documentclass[10pt,twocolumn,letterpaper]{article}

\usepackage[pagenumbers]{cvpr} 

\input{preamble}
\definecolor{cvprblue}{rgb}{0.21,0.49,0.74}
\usepackage[pagebackref,breaklinks,colorlinks,allcolors=cvprblue]{hyperref}
\usepackage{multirow}
\usepackage{stfloats}
\def\paperID{6736} 
\def\confName{CVPR}
\def\confYear{2026}

\title{EditEmoTalk: Controllable Speech-Driven 3D Facial Animation with Continuous Expression Editing}

\author{
  Diqiong Jiang\textsuperscript{1},
  Kai Zhu\textsuperscript{1},
  Dan Song\textsuperscript{2},
  Jian Chang\textsuperscript{3},
  Chenglizhao Chen\textsuperscript{1}\textsuperscript{*},
  Zhenyu Wu\textsuperscript{4}\\
  \textsuperscript{1}China University of Petroleum,
  \textsuperscript{2}Tianjin University,
  \textsuperscript{3}Bournemouth University,\\
  \textsuperscript{4}Southwest Jiaotong University\\
  \normalsize
}

\begin{document}
\twocolumn[{
\renewcommand\twocolumn[1][]{#1}
\maketitle
\begin{center}
    \includegraphics[width=\textwidth,,trim={2cm 6.5cm 2.4cm 5.3cm}, clip]{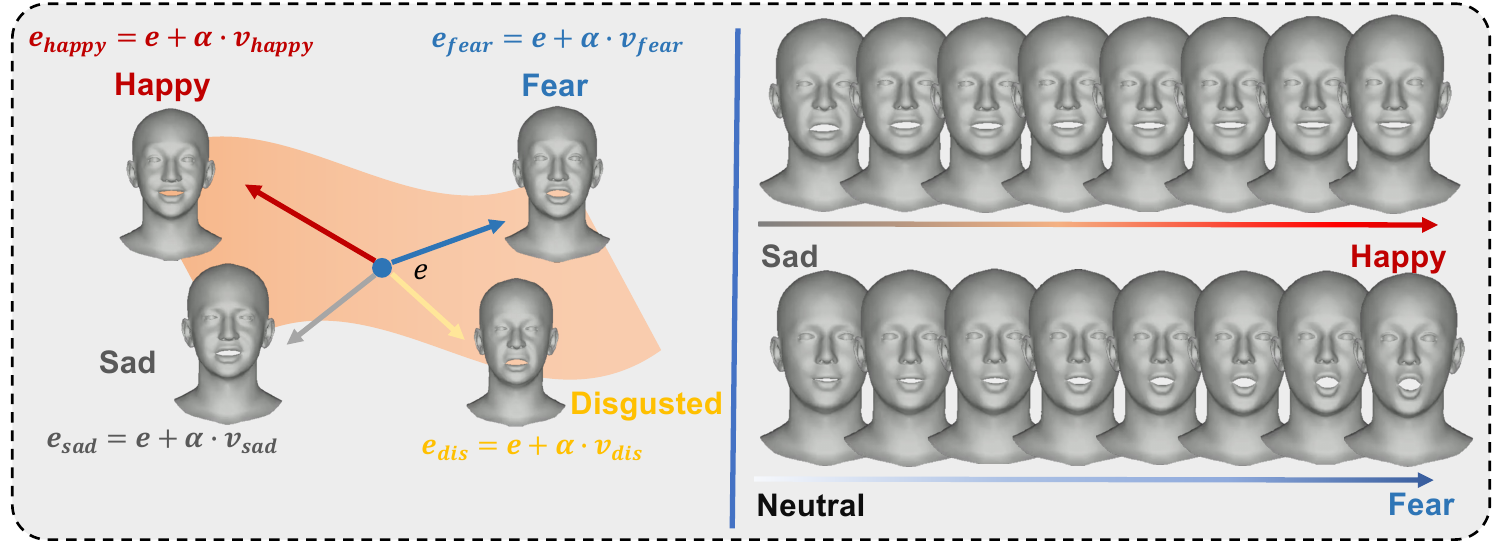}
    \captionof{figure}{
        \textbf{EditEmoTalk: Continuous Emotion Editing for Speech-Driven 3D Facial Animation.}
        (Left) Emotion editing in the learned expression manifold. Each point represents an expression embedding, where smooth interpolation across the manifold enables natural emotional transitions. 
        (Right) Continuous 3D facial mesh animation results generated by EditEmoTalk. The top row shows a smooth transition from \textit{sad $\rightarrow$ happy}, and the bottom row demonstrates \textit{neutral $\rightarrow$ fear}, illustrating fine-grained emotional control while maintaining natural lip synchronization.
    }
    \label{fig:teaser}
\end{center}
\vspace{4mm}
}]
\input{sec/0_abstract}    
\input{sec/1_intro}
\input{sec/2_formatting}
\input{sec/3_finalcopy}

\input{sec/4_experiment}

{
    \small
    \bibliographystyle{ieeenat_fullname}
    \bibliography{main}
}


\end{document}

%% file: sec/0_abstract.tex

\begin{abstract}
Speech-driven 3D facial animation aims to generate realistic and expressive facial motions directly from audio. While recent methods achieve high-quality lip synchronization, they often rely on discrete emotion categories, limiting continuous and fine-grained emotional control. We present EditEmoTalk, a controllable speech-driven 3D facial animation framework with continuous emotion editing. The key idea is a boundary-aware semantic embedding that learns the normal directions of inter-emotion decision boundaries, enabling a continuous expression manifold for smooth emotion manipulation. Moreover, we introduce an emotional consistency loss that enforces semantic alignment between the generated motion dynamics and the target emotion embedding through a mapping network, ensuring faithful emotional expression. Extensive experiments demonstrate that EditEmoTalk achieves superior controllability, expressiveness, and generalization while maintaining accurate lip synchronization. Code and pretrained models will be released.
\end{abstract}

%% file: sec/1_intro.tex
\section{Introduction}
\label{sec:intro}

Generating natural and expressive 3D facial animation from speech is a long-standing challenge in computer graphics, computer vision, and human–computer interaction. Speech-driven animation offers a cost-effective alternative to manual production and enables scalable deployment of digital humans in applications such as virtual assistants, gaming, film production, and immersive communication. Despite significant advances in deep learning-based methods~\cite{jiang2024audio, thambiraja2023imitator, chai2022personalized, fan2022faceformer, xing2023codetalker}, current approaches remain limited in their ability to capture and control emotional expressiveness.

Most existing techniques~\cite{jafari2024jambatalk, ji2024realtalk, stan2023facediffuser} directly map speech to facial motion, often resulting in rigid and emotionally neutral expressions. Expression-conditioned methods~\cite{peng2023emotalk, danvevcek2023emotional, feng2025faceedittalker, liang2024cstalk} incorporate emotion cues but typically rely on discrete emotion categories (e.g., “happy,” “sad”), which constrain smooth interpolation and continuous control. Consequently, they struggle to produce nuanced emotional transitions or fine-grained intensity modulation—key ingredients for achieving believable and interactive digital humans.

To overcome these limitations, we present \textbf{EditEmoTalk}, a controllable speech-driven 3D facial animation framework that supports \textit{continuous emotion editing}. Instead of relying on categorical emotion embeddings, our method learns the normals to inter-emotion decision boundaries, thereby structuring a continuous \textit{expression manifold} around perceptually meaningful emotional transitions. Editing along these boundary-aware directions ensures that the generated trajectories remain within semantically valid and emotionally distinct regions, enabling smooth, interpretable, and emotion-consistent facial editing while preserving high-fidelity speech synchronization.

To further ensure semantic alignment between the edited emotion embeddings and the resulting motion dynamics, we introduce an \textit{emotional consistency loss}. This objective enforces correspondence between the edited emotion vector and the generated facial motion through a learned mapping network, guaranteeing that the output faithfully reflects the intended affective semantics. In addition, a \textit{dual-train strategy} alternates stochastically between speech-driven and emotion-edited inputs during training. This design allows the model to robustly maintain natural lip-sync when no editing is applied, while enabling precise and continuous emotion manipulation when edits are introduced.

By reformulating emotion control as continuous traversal on a learned expression manifold, \textbf{EditEmoTalk} unifies discrete emotion classification and continuous emotion editing within a single framework.

\noindent \textbf{Our main contributions are as follows:}
\begin{itemize}
\item \textbf{EditEmoTalk:} A controllable speech-driven 3D facial animation framework that achieves \textit{continuous} and \textit{fine-grained} emotional control, bridging discrete emotion recognition and continuous expression editing.
\item \textbf{Boundary-aware semantic embedding:} A novel representation that learns inter-emotion boundary normals to construct a continuous \textit{expression manifold}, enabling smooth and interpretable emotion transitions.
\item \textbf{Emotional consistency loss and dual-train strategy:} A synergistic design that aligns motion dynamics with edited emotion embeddings while preserving natural lip synchronization and semantic coherence.
\end{itemize}

%% file: sec/2_formatting.tex
\section{Related Work}
\label{sec:RelatedWork}

\subsection{Speech-Driven 3D Facial Animation}

Speech-driven facial animation has long been a central problem in computer vision and computer graphics. Early methods relied on linguistic analysis, mapping phonemes to visemes using hand-crafted coarticulation rules~\cite{thalmann1993models}. Statistical approaches such as Hidden Markov Models (HMMs)~\cite{brand1999voice} and Gaussian Process Latent Variable Models (GPLVMs)~\cite{deena2009speech,deena2013visual} were later introduced to capture temporal dependencies from motion trajectories.

With the advent of deep learning, CNN- and RNN-based models began predicting facial movements directly from audio~\cite{suwajanakorn2017synthesizing,karras2017audio,taylor2017deep}, achieving more accurate lip synchronization than traditional methods. Recent transformer- and diffusion-based approaches~\cite{fan2022faceformer,thambiraja2023imitator,peng2023emotalk,danvevcek2023emotional} further enhance temporal coherence and expressiveness. Mesh-based representations~\cite{lu2021live,tian2024emo,xu2024vasa} provide higher geometric fidelity compared to image-based methods, enabling more realistic facial motion. Despite these advances, most existing methods focus primarily on lower-face movements and accurate lip synchronization, often overlooking upper-face dynamics such as eyebrow and eye motions, which are critical for conveying emotion. In addition, emotion is typically represented using discrete categories or global style codes, limiting smooth interpolation and fine-grained control.

\subsection{Facial Expression Editing and Control}

Facial expression editing and control have been widely explored in both 2D and 3D domains. In images and video, GAN- and diffusion-based methods achieve photorealistic manipulation through latent space traversal or semantic conditioning~\cite{huang2023collaborative,jiang2023styleipsb,melnik2024face,patashnik2021styleclip}. In 3D modeling, expression control is commonly realized via blendshape interpolation, latent disentanglement, or parametric models that separate identity and expression factors~\cite{blanz1999morphable,li2017learning,jiang2019disentangled}, supporting interpolation, intensity scaling, and cross-subject transfer.

Several recent works attempt to incorporate emotion into speech-driven animation~\cite{han2025pestalk,peng2023emotalk,danvevcek2023emotional,feng2025faceedittalker}, but they largely rely on discrete emotion categories or low-dimensional style embeddings. This restricts fine-grained control and prevents continuous transitions between emotional states, resulting in animations with limited expressiveness.

In contrast, our work introduces a unified framework for continuous, emotion-aware control in speech-driven 3D facial animation. By constructing a continuous expression manifold via a boundary-aware semantic embedding and enforcing temporal emotional consistency through motion alignment, EditEmoTalk enables fine-grained, smoothly interpolatable emotional control across both lower- and upper-face movements, bridging the gap between realistic speech-driven dynamics and flexible, expressive animation.

%% file: sec/3_finalcopy.tex
\section{Method}

\begin{figure*}[htbp]
\centering
\includegraphics[width=\textwidth,trim={0cm 3cm 0.0cm 2cm}, clip]{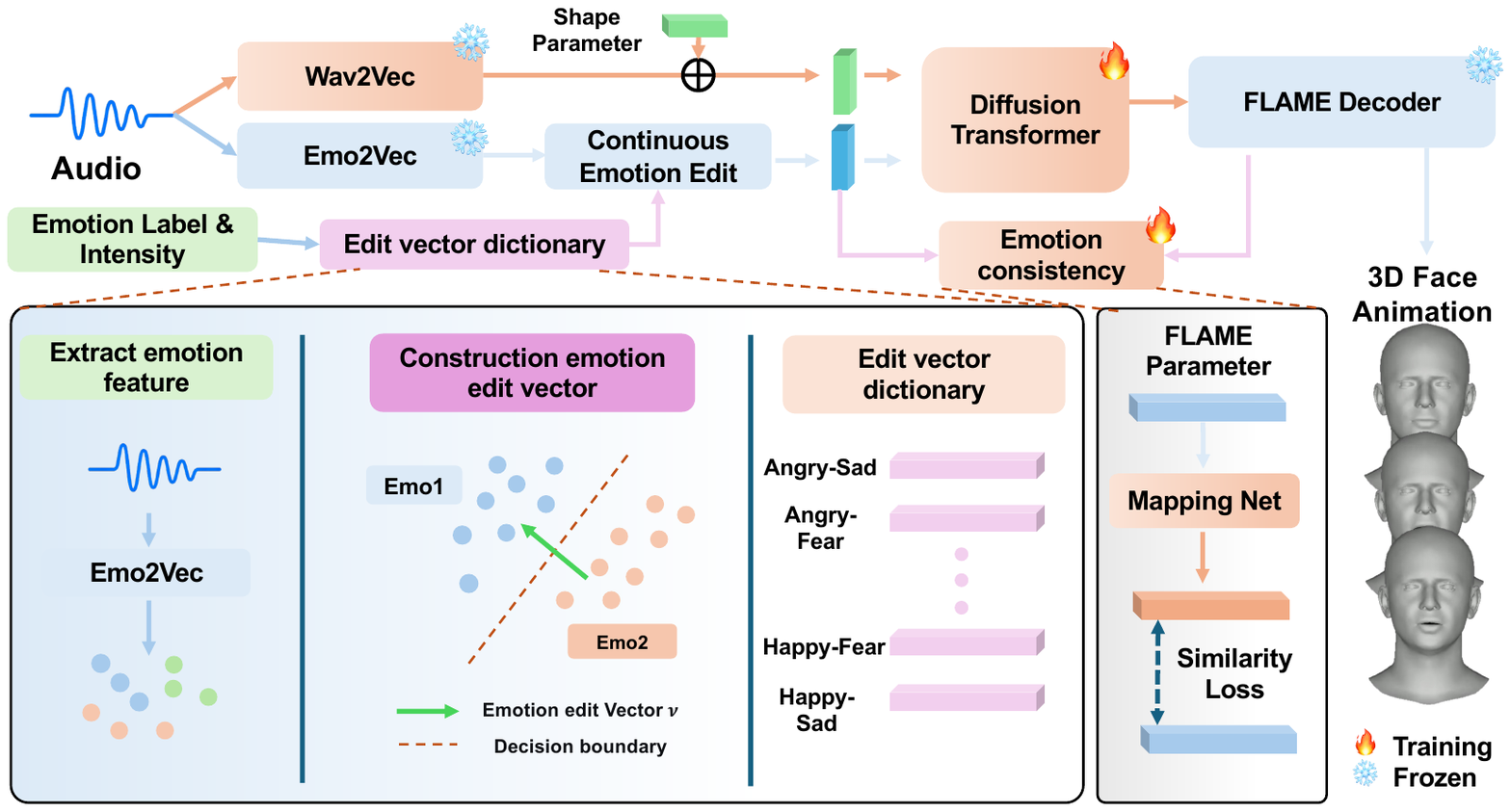}
\caption{\textbf{Overview of our controllable 3D facial animation framework.} Our model generates 3D facial motion from speech audio with continuous emotion editing capability. The pipeline begins by extracting speech features (HuBERT) and emotion cues (emo2vec). The key to our continuous editing lies in the expression conditioning of the diffusion process: expression vectors from our boundary-aware Edit Vector Dictionary are combined with audio features and FLAME parameters to form the conditioning signal for the Diffusion Transformer. This conditions the generative denoising process, enabling precise steering of the output expression—where adjusting the expression vector directly modulates the emotional style of the generated mesh sequence while maintaining lip synchronization. The Mesh Decoder then produces the final 3D facial animations.}
\label{fig:overview}
\end{figure*}

As illustrated in Figure~\ref{fig:overview}, \textbf{EditEmoTalk} consists of four main components that collaboratively produce controllable and emotionally expressive 3D facial animations. The pipeline begins with a speech feature encoder that extracts discriminative acoustic representations from raw audio, which are fused with expression embeddings retrieved from the \textbf{Edit Vector Dictionary} to provide emotion conditioning. Finally, the \textbf{Continuous Expression Editing Module} enables modulation of expression intensity and style guided by the continuous emotion manifold, while an emotional consistency loss enforces semantic alignment between emotion embeddings and motion dynamics.

\subsection{Speech Feature Encoder}

To capture both linguistic and affective aspects of speech, we adopt a dual-branch encoder that decomposes the raw waveform into \textit{content} and \textit{emotion} components.

\begin{itemize}
    \item \textbf{Content Encoder:} A pre-trained HuBERT~\cite{hsu2021hubert} model serves as the content encoder $\mathcal{E}_c$, extracting phoneme-level representations $A = \mathcal{E}_c(X)$ from the input waveform $X$.

    \item \textbf{Emotion Encoder:} We employ emotion2vec~\cite{ma2023emotion2vec} as the emotion encoder $\mathcal{E}_e$ to obtain a compact embedding $e = \mathcal{E}_e(X)$ that captures the utterance’s overall emotional state. This embedding provides the basis for continuous emotion editing in our framework.
\end{itemize}

\subsection{Continuous Expression Editing Module}

To enable fine-grained and continuous emotion control, we introduce a \textit{Continuous Expression Editing Module} that operates in a learned emotion feature space. Unlike discrete label-based methods, it models the geometric structure of the emotion space, allowing smooth and interpretable expression manipulation.

\noindent \textbf{Emotion Space.}
Let $E = \{e_1, e_2, \ldots, e_N\}$ denote emotion embeddings extracted from a labeled dataset, where each $e_i \in \mathbb{R}^d$ corresponds to an audio segment with label $y_i \in \{1, \ldots, K\}$. These embeddings form clusters reflecting the semantic topology of human emotions. Our goal is to uncover the geometric relations between clusters for continuous editing.

\noindent \textbf{Boundary-Aware Semantic Embedding.}
The key insight of our editing strategy is that the most semantically meaningful directions for emotion transition are encoded in the decision boundaries between emotion classes. Intuitively, the normal vector of a decision boundary points in the direction of steepest ascent away from one emotion class and toward another. We therefore estimate a local separating hyperplane for each emotion $k$ by training a multi-class linear classifier on the emotion embeddings $E$. The hyperplane is given by:
\[
f_k(e) = \mathbf{w}_k^T e + b_k = 0,
\]
where $\mathbf{w}_k$ is the normal direction of the decision boundary. Instead of serving as a classifier, these boundaries capture semantic transitions between neighboring emotions, forming geometry-aware representations that encode relative emotional variations.

\noindent \textbf{Continuous Emotion Manifold.}
Normalized boundary normals define principal editing directions:
\[
\mathbf{v}_k = \frac{\mathbf{w}_k}{\|\mathbf{w}_k\|_2},
\]
each corresponding to a tangent direction on the emotion manifold. Collectively, these directions span a locally linear subspace approximating a \textit{continuous emotion manifold}, enabling smooth transitions across emotional states.

\noindent \textbf{Continuous Control.}
Given an input embedding $e_{orig}$, emotion intensity is adjusted by a scalar $\alpha$:
\[
e_{edited} = e_{orig} + \alpha \cdot \mathbf{v}_k,
\]
or, for multi-emotion editing,
\[
e_{edited} = e_{orig} + \sum_{k=1}^K \alpha_k \cdot \mathbf{v}_k.
\]
This supports intuitive continuous transitions such as from neutral to smiling or from calm to excited. The resulting edited embedding $e_{edited}$ is then fed as a conditioning signal to the cross-attention layers of our Diffusion Transformer, thereby steering the generative process to output meshes consistent with the target expression.

\noindent \textbf{Implementation.}
The decision boundaries are computed offline, and the normalized direction vectors $\{\mathbf{v}_1, \ldots, \mathbf{v}_K\}$ are stored in an \textbf{Edit Vector Dictionary} for real-time inference. This design:  
(1) requires only categorical labels instead of intensity annotations;  
(2) offers semantically interpretable, geometry-consistent editing; and  
(3) preserves speech-content correlation by modifying only emotion-related features.

\subsection{Diffusion Transformer and Mesh Decoder}

We employ a conditional diffusion model to generate 3D facial motion from speech. Our Diffusion Transformer (DiT) backbone~\cite{peebles2023scalable} denoises FLAME parameter sequences $\mathbf{\Theta}_t \in \mathbb{R}^{T \times D}$ conditioned on speech features ($A$) from HuBERT, emotion embeddings ($e$) from our Edit Vector Dictionary, and identity coefficients ($\beta$). The denoising process $\mathbf{\Theta}_{pred} = \text{DiTBlock}(\mathbf{\Theta}_t, t, A, e, \beta)$ ensures emotional consistency while preserving lip synchronization.

Final denoised parameters $\mathbf{\Theta}_0 = \{\beta, \psi, \theta\}$ are decoded through the FLAME model $\mathcal{M} = \mathcal{G}_{FLAME}(\beta, \psi, \theta)$ to produce the output 3D mesh sequence, maintaining temporal coherence and mesh quality.

\subsection{Training Objectives}

We employ a multi-level objective function to ensure high-quality 3D facial animation generation, encompassing parameter reconstruction, geometric fidelity, temporal coherence, and emotional consistency.

\subsubsection{Reconstruction Loss}

A masked mean squared error (MSE) loss is applied to enforce parameter-level reconstruction accuracy between predicted and ground-truth FLAME motions:
\[
\mathcal{L}_{recon} = \frac{1}{T D}  \sum_{t=1}^{T} \|\mathbf{P}_{t} - \mathbf{P}_{t}^{gt}\|_2^2,
\]
where $\mathbf{P} \in \mathbb{R}^{T \times D}$ denotes the predicted motion parameters, $\mathbf{P}^{gt}$ the corresponding ground truth, $T$ the sequence length, and $D$ the parameter dimension.

\subsubsection{Mesh-Level Geometric Losses}

To ensure geometric fidelity of the generated 3D meshes, we define vertex-level geometric losses computed on the decoded FLAME outputs:

\begin{itemize}
    \item \textbf{Mesh Reconstruction Loss:} Enforces overall geometric accuracy by minimizing vertex-wise Euclidean distances:
    \[
    \mathcal{L}_{mesh} = \frac{1}{T V} \sum_{t=1}^{T} \|\mathbf{V}_t - \mathbf{V}_t^{gt}\|_2^2,
    \]
    where $\mathbf{V}_t \in \mathbb{R}^{V \times 3}$ and $\mathbf{V}_t^{gt}$ denote predicted and ground-truth vertex positions, respectively, and $V$ is the vertex count.
    
    \item \textbf{Surface Normal Loss:} Preserves local geometric smoothness by penalizing deviations in vertex normals:
    \[
    \mathcal{L}_{normal} = \frac{1}{T V} \sum_{t=1}^{T} \|\mathbf{N}_t - \mathbf{N}_t^{gt}\|_2^2,
    \]
    where $\mathbf{N}_t$ and $\mathbf{N}_t^{gt}$ are predicted and ground-truth vertex normals.
\end{itemize}

\subsubsection{Temporal Coherence Losses}

Temporal consistency is encouraged through first- and second-order smoothness constraints:

\begin{itemize}
    \item \textbf{Mesh Velocity Loss:} Maintains first-order temporal continuity:
    \[
    \mathcal{L}_{vel} = \frac{1}{(T-1)V} \sum_{t=1}^{T-1} \|\Delta\mathbf{V}_t - \Delta\mathbf{V}_t^{gt}\|_2^2,
    \]
    where $\Delta\mathbf{V}_t = \mathbf{V}_{t+1} - \mathbf{V}_t$.
    
    \item \textbf{Mesh Acceleration Loss:} Enforces second-order smoothness:
    \[
    \mathcal{L}_{acc} = \frac{1}{(T-2)V} \sum_{t=1}^{T-2} \|\Delta^2\mathbf{V}_t - \Delta^2\mathbf{V}_t^{gt}\|_2^2,
    \]
    where $\Delta^2\mathbf{V}_t = \mathbf{V}_{t+2} - 2\mathbf{V}_{t+1} + \mathbf{V}_t$.
\end{itemize}

\subsubsection{Emotion Consistency Loss}
We introduce a lightweight \textbf{Mapping Network} with two fully connected layers and GELU activations to project predicted expression parameters into the space of the \textbf{emotional consistency loss}, enforcing alignment between edited embeddings and generated motion. The \textbf{emotion similarity loss} is defined via cosine distance:
\[
\mathcal{L}_{emo} = 1 - \frac{\mathbf{e} \cdot \mathbf{e}^{gt}}{\|\mathbf{e}\|_2 \, \|\mathbf{e}^{gt}\|_2},
\]
where $\mathbf{e}$ and $\mathbf{e}^{gt}$ are the predicted and target emotion embeddings.

\subsubsection{Overall Objective and Training Strategy}

The final objective integrates all components with empirically tuned weights:
\[
\begin{aligned}
\mathcal{L}_{total} = & \ \lambda_{recon} \mathcal{L}_{recon} + \lambda_{mesh} \mathcal{L}_{mesh} + \lambda_{normal} \mathcal{L}_{normal} \\
& + \lambda_{vel} \mathcal{L}_{vel} + \lambda_{acc} \mathcal{L}_{acc} + \lambda_{emo} \mathcal{L}_{emo}.
\end{aligned}
\]

To effectively train the model for both accurate speech-driven generation and continuous emotion editing, we adopt a stochastic training strategy: with 50\% probability, the input is the emotion embedding extracted from audio and the model is optimized with the full objective $\mathcal{L}_{total}$, while in the remaining 50\% of cases, an edited emotion embedding from the continuous expression manifold is used and only the emotional consistency loss
\[
\mathcal{L}_{edit} = \lambda_{emo} \mathcal{L}_{emo}
\]
is applied. This strategy encourages both faithful speech-driven animation and precise, continuous emotion control.

This mixed strategy ensures that the model learns to generate realistic and temporally coherent facial animations while also being capable of precise and continuous manipulation of emotional expressions.

%% file: sec/4_experiment.tex
\section{Experiments}

We conduct a thorough evaluation of the proposed \textbf{EditEmoTalk} framework to verify its effectiveness in terms of generation realism, lip-sync accuracy and expression controllability. We further perform ablation studies to justify our key design choices.

\subsection{Datasets and Preprocessing}

Our model is trained and evaluated on four publicly available audio-visual datasets to ensure diversity in speakers, languages, and emotional expressions:

\textbf{MEAD}~\cite{kaisiyuan2020mead}: A high-quality, multi-view emotional audio-visual dataset featuring 60 actors performing eight emotions at three intensity levels, captured under strictly controlled conditions with seven synchronized cameras. It serves as our primary dataset for controllable emotion modeling. \textbf{CREMA-D}~\cite{cao2014crema}: Contains 7,442 clips from 91 diverse actors performing 12 sentences with six emotions at four intensity levels. The dataset includes large-scale crowd-sourced emotion ratings, enabling robust evaluation of emotion generalization. \textbf{RAVDESS}~\cite{livingstone2018ryerson}: Comprises 7,356 recordings from 24 professional actors performing speech and song with eight emotions at two intensity levels, available in audio-only, video-only, and audio-visual modalities, supporting multimodal emotion analysis. \textbf{HDTF}~\cite{zhang2021flow}: Consists of about 16 hours of high-resolution talking-face videos collected from YouTube. As an in-the-wild dataset, it provides diverse poses, lighting, and backgrounds, serving as a challenging benchmark for real-world generalization.

\textbf{Preprocessing:} Audio is resampled to 16 kHz. We employ an off-the-shelf 3D face reconstruction method~\cite{retsinas20243d} to obtain ground-truth 3D mesh sequences as training targets, and facial motion is represented as vertex displacements from a neutral frame.

\subsection{Evaluation Metrics}

We evaluate EditEmoTalk using quantitative and qualitative metrics. Quantitatively, \textit{Vertex Error (VE)} measures the average L2 distance (in mm) between predicted and ground-truth vertices, reflecting overall reconstruction accuracy. \textit{Lip Vertex Error (LVE)}~\cite{richard2021meshtalk} evaluates lip-sync precision, \textit{Mouth Opening Deviation (MOD)}~\cite{cudeiro2019capture} measures articulation amplitude, and \textit{Facial Dynamics Deviation (FDD)}~\cite{xing2023codetalker} assesses temporal motion smoothness by comparing vertex velocities. 

In addition, we introduce \(\Delta{\text{CH}}\), which evaluates the preservation of emotion clustering structure in the generated motions using the Calinski–Harabasz (CH) index. 
The CH index measures the ratio of between-cluster dispersion to within-cluster dispersion in the \textit{FLAME expression parameter space}. 
Smaller \(\Delta{\text{CH}}\) values indicate that the generated motions maintain emotion clusters that are more compact internally and well-separated from other emotions, closely matching the ground-truth emotional manifold.

Unlike prior works, we \textbf{do not report emotion classification accuracy}. A higher classification score does not necessarily reflect more expressive or realistic animation; instead, enforcing discrete emotion categories often reduces \textbf{temporal coherence} and increases LVE, as rigid class boundaries conflict with the \textbf{continuous dynamics of human emotion}.

Qualitatively, a user study reports \textit{Mean Opinion Score (MOS)} on a 5-point scale for naturalness, lip-sync, and emotional authenticity, comparing EditEmoTalk to baseline methods.

\subsection{Baselines}

EditEmoTalk is compared with several state-of-the-art 3D talking head generation methods. \textbf{DiffPoseTalk}~\cite{sun2024diffposetalk} uses pose-level diffusion for expressive motion with temporal smoothness. \textbf{Emote}~\cite{danvevcek2023emotional} employs discrete emotion embeddings for controllable expressions. \textbf{EmoTalk}~\cite{peng2023emotalk} introduces an Emotion Disentangling Encoder (EDE) to separate emotion and content via cross-reconstruction, enabling controllable expression synthesis while preserving lip-sync. \textbf{DeepTalk}~\cite{kim2025deeptalk} trains a Dynamic Emotion Embedding (DEE) module with probabilistic contrastive learning to jointly model speech and facial motion, capturing nuanced emotional variations.

\subsection{Implementation Details}

Our model is implemented in PyTorch and trained with PyTorch Lightning using the AdamW optimizer
($\text{lr}=1\times10^{-4}$, weight decay $1\times10^{-5}$) and a cosine annealing scheduler.
Training is performed for 200 epochs with a batch size of~8 on a single RTX~4090 GPU.
The model contains 446M parameters. During training, emotion editing is randomly enabled to promote disentangled control.
At inference, an emotion vector~$\mathbf{e}$ (one-hot or continuous) allows flexible emotion manipulation. \textbf{Inference speed.}
The model runs at 0.035\,ms per frame (to FLAME parameters), achieving real-time performance.

\begin{table*}[t]
\centering
\small
\setlength{\tabcolsep}{2pt}
\caption{Cross-dataset generalization performance comparison. 
\(\Delta{\text{CH}}\) is omitted for HDTF as it lacks consistent emotion labels required for clustering evaluation.
Since EmoTalk does not adopt a FLAME-based facial topology, certain metrics (e.g., FDD and $\Delta \textbf{CH}$) are not directly comparable and thus omitted.}
\label{tab:cross_dataset_ch}
\begin{tabular*}{\textwidth}{@{\extracolsep{\fill}}l l c c c c c}
\hline
\textbf{Dataset} & \textbf{Method} & \textbf{VE (mm)} $\downarrow$ & \textbf{LVE (mm)} $\downarrow$ & \textbf{MOD (mm)} $\downarrow$ & \textbf{FDD $(10^{-5}$ mm/s)} $\downarrow$ & $\Delta{\textbf{CH}}$ $\downarrow$ \\
\hline
\multirow{5}{*}{\textbf{CREMA-D}} 
& DeepTalk & 2.0733 & 11.4756 & 1.7794 & 42.8523 & 4.7572 \\
& DiffPoseTalk & 1.8965 & 12.3115 & 2.4100 & 31.2141 & 0.8483 \\
& EmoTalk & 2.4353 & 12.4324 & 1.8323 & -- & 0.9110 \\
& Emote & 1.8902 & 9.7072 & 1.7654 & 64.4193 & 62.8581 \\
& \textbf{Ours} & \textbf{0.5553} & \textbf{4.8682} & \textbf{0.9318} & \textbf{30.8475} & \textbf{0.0865} \\
\hline
\multirow{5}{*}{\textbf{RAVDESS}} 
& DeepTalk & 2.2248 & 13.2099 & 2.3545 & 33.4368 & 2.2636 \\
& DiffPoseTalk & 1.6545 & 9.4647 & 1.5972 & 39.3294 & 0.8540 \\
& EmoTalk & 2.3432 & 10.3224 & 2.3324 & -- & 0.8539 \\
& Emote & 2.0747 & 10.9277 & 2.3453 & 55.2231 & 17.708 \\
& \textbf{Ours} & \textbf{0.6940} & \textbf{6.4486} & \textbf{1.1563} & \textbf{33.3030} & \textbf{0.0942} \\
\hline
\multirow{5}{*}{\textbf{MEAD}} 
& DeepTalk & 2.0430 & 11.5721 & 2.0534 & 28.0587 & 0.6194 \\
& DiffPoseTalk & 1.6939 & 10.6261 & 1.7888 & 31.6314 & 0.9655 \\
& EmoTalk & 2.5432 & 11.2343 & 1.7943 & -- & 0.9888 \\
& Emote & 1.8046 & 8.6625 & 1.5026 & 49.3305 & 35.855 \\
& \textbf{Ours} & \textbf{0.5340} & \textbf{4.7738} & \textbf{0.9179} & \textbf{25.3608} & \textbf{0.0623} \\
\hline
\multirow{5}{*}{\textbf{HDTF}} 
& DeepTalk & 2.0551 & 13.6581 & 2.6189 & 69.7752 & -- \\
& DiffPoseTalk & 1.5078 & 8.5308 & 1.1930 & 87.2130 & -- \\
& EmoTalk & 1.9344 & 11.2342 & 2.0892 & -- & -- \\
& Emote & 1.6248 & 9.8394 & 2.2399 & 89.8650 & -- \\
& \textbf{Ours} & \textbf{0.8084} & \textbf{7.8892} & \textbf{1.7177} & \textbf{86.8194} & -- \\
\hline
\end{tabular*}
\end{table*}

\subsection{Results and Analysis}

\subsubsection{Quantitative Results}
\noindent
\textbf{Cross-Dataset Analysis.}  
Table~\ref{tab:cross_dataset_ch} presents a comprehensive cross-dataset comparison, incorporating the proposed $\Delta_{\text{CH}}$ metric to evaluate emotion consistency.  
Across all datasets, EditEmoTalk achieves the lowest VE and LVE, indicating superior geometric accuracy and lip-sync precision. It also maintains competitive MOD and FDD values, reflecting stable articulation amplitude and smooth temporal motion.  

Beyond motion quality, EditEmoTalk demonstrates the smallest $\Delta{\text{CH}}$ across CREMA-D, RAVDESS, and MEAD, suggesting that its generated facial motions preserve emotion clustering structures closest to the ground truth.  
In contrast, methods such as DeepTalk and DiffPoseTalk exhibit larger $\Delta{\text{CH}}$ values, implying weaker emotional separation and less coherent affective representation.  
Notably, Emote, despite producing visually expressive motion, shows extremely high \(\Delta{\text{CH}}\) values. We attribute this to its reliance on discrete emotion embeddings that enforce rigid class boundaries, which conflicts with the continuous nature of emotional expressions and results in artificial clustering patterns that diverge from the natural emotional manifold.

\begin{table}[h]
\centering
\caption{User study results (Mean Opinion Scores, 1–5). Our \textbf{EditEmoTalk} achieves the highest perceptual quality across all evaluation aspects.}
\label{tab:user_study}
\setlength{\tabcolsep}{4.2pt}
\begin{tabular}{lccc}
\hline
\textbf{Method} & \textbf{Naturalness} $\uparrow$ & \textbf{Lip-sync} $\uparrow$ & \textbf{Emotion} $\uparrow$ \\
\hline
DiffPoseTalk & 3.82 & \textbf{4.15} & 3.05 \\
DeepTalk & 3.56 & 3.47 & 3.29 \\
EmoTalk & 3.21 & 3.88 & 3.67 \\
Emote & 4.07 & 3.74 & 3.93 \\
\textbf{Ours} & \textbf{4.46} & 4.07 & \textbf{4.49} \\
\hline
\end{tabular}
\end{table}

\subsubsection{Qualitative Results}
We further perform qualitative evaluations to visually assess the perceptual quality of the generated facial animations. In this section, we focus on two key aspects: \textbf{expression richness} and \textbf{controllability analysis}. The first aspect examines how well different methods reproduce diverse and natural facial expressions corresponding to emotional speech, while the second evaluates the ability of the model to control and edit emotional intensity or category in a consistent and realistic manner.

\begin{figure*}[t]
\centering
\includegraphics[width=0.99\linewidth,trim={1.5cm 4cm 2.5cm 4cm},clip]{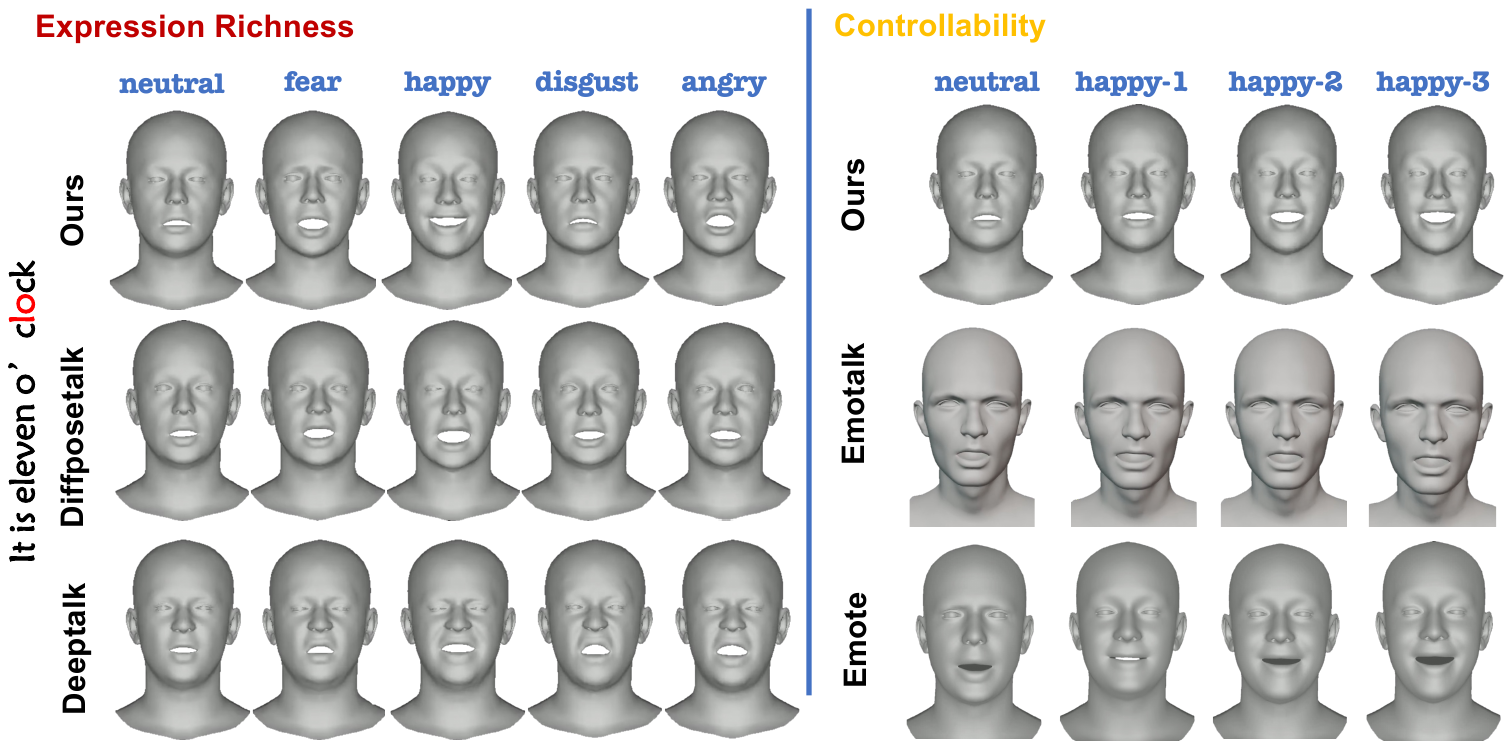}
\caption{
\textbf{Left:} Comparison of expression richness across different methods. Each method generates facial animations driven by the same utterance “\textit{It is eleven o’clock},” with emotional emphasis on the syllable “\textit{lo}.” Our method produces richer emotional expressiveness and smoother facial dynamics, particularly in the depiction of nuanced expressions such as \textit{angry}, \textit{happy}, \textit{neutral}, \textit{fear}, and \textit{disgust}. 
\textbf{Right:} Controllability analysis showing natural expressions and three levels of edited smiling intensity.
}
\label{fig:expression_richness}
\end{figure*}

\begin{table}[h]
\centering
\caption{Ablation study on MEAD validation set. Both the dual-train strategy and emotional consistency loss contribute to improved facial animation quality. LVE and MOD are in mm, FDD is in $10^{-5}$ mm/s.}
\label{tab:ablation}
\setlength{\tabcolsep}{6pt}
\begin{tabular}{lccc}
\hline
\textbf{Model Variant} & \textbf{LVE} $\downarrow$ & \textbf{MOD} $\downarrow$ & \textbf{FDD} $\downarrow$ \\
\hline
Full Model & \textbf{4.7738} & \textbf{0.9179} & \textbf{28.3608} \\
w/o Dual-Train & 5.1024 & 0.9451 & 29.6286 \\
w/o $\mathcal{L}_{emo}$ & 4.9261 & 0.9324 & 30.7245 \\
\hline
\end{tabular}
\end{table}

\subsubsection{Expression Richness}
To further evaluate the expressiveness of generated animations, we visualize the results of different methods in Figure~\ref{fig:expression_richness}. The input speech is “\textit{It is eleven o’clock},” where the emphasis naturally falls on the syllable “\textit{lo},” accompanied by five distinct emotions: \textit{angry}, \textit{happy}, \textit{neutral}, \textit{fear}, and \textit{disgust}.
Our proposed \textbf{EditEmoTalk} demonstrates superior \textit{expression richness} and \textit{facial dynamics}, producing subtle, temporally coherent expressions that align closely with both the speech rhythm and emotional cues.
Although \textbf{DiffPoseTalk} adopts a diffusion-based architecture similar to ours, it lacks the dual-channel structure and emotion-consistency loss incorporated in \textbf{EditEmoTalk}. As a result, its outputs exhibit limited emotional variation, often remaining close to a neutral or weakly expressive state.
\textbf{DeepTalk} shows clear intensity for \textit{angry} and \textit{disgust} emotions; however, its generated results tend to share similar facial tension patterns across different emotions, leading to poor separation and reduced emotional diversity.
In contrast, our method achieves more distinguishable and naturally evolving emotional expressions across categories, effectively enhancing both emotional realism and visual appeal in speech-driven 3D facial animation.


\subsubsection{Controllability Analysis}

A key advantage of \textbf{EditEmoTalk} lies in its explicit and fine-grained emotion controllability.
By manipulating the expression control vector $\mathbf{e}$, our model can render the same speech input into different emotional states without retraining.

We further conduct a qualitative comparison with \textbf{EmoTalk} and \textbf{Emote} to evaluate controllability and visual expressiveness. As shown in Figure~\ref{fig:expression_richness}, EmoTalk exhibits limited editing capability—its generated expressions remain nearly unchanged across different emotion settings. Emote achieves visible variations between emotional states; however, the emotional intensity and facial dynamics are less pronounced and sometimes inconsistent. In contrast, our \textbf{EditEmoTalk} produces clearly distinguishable emotional expressions with smooth transitions and consistent lip synchronization, effectively disentangling emotion-related facial deformation from speech-driven motion.

This controllable synthesis demonstrates the potential of \textbf{EditEmoTalk} for expressive and interactive digital human applications, such as virtual presenters, digital assistants, and emotionally aware avatars.

\subsubsection{User Study}

To further evaluate the perceptual quality and emotional fidelity of our generated facial animations, we conducted a comprehensive user study focusing on three key aspects: naturalness, lip synchronization accuracy, and emotional authenticity. A total of 30 participants (15 male and 15 female, aged 20–35) were invited to evaluate rendered videos from five methods: DiffPoseTalk, DeepTalk, EmoTalk, Emote, and our proposed EditEmoTalk. Each participant viewed 20 randomly selected samples across various emotional categories, with both the order of methods and emotion types randomized to eliminate potential bias. Participants rated each clip using a 5-point Mean Opinion Score (MOS) scale (1 = very poor, 5 = excellent) in terms of overall realism and motion smoothness for naturalness, the alignment between lip movement and speech for lip-sync accuracy, and the clarity and realism of expressed emotions for emotional authenticity.

As presented in Table~\ref{tab:user_study}, EditEmoTalk achieves the highest overall Mean Opinion Scores (MOS) across all perceptual aspects.
While diffusion-based \textbf{DiffPoseTalk} exhibits strong lip synchronization performance due to its pose-level modeling, it lacks expressive diversity, resulting in monotonous facial dynamics.
\textbf{DeepTalk} produces coherent but relatively rigid motion, and \textbf{EmoTalk} offers emotion control yet fails to maintain smooth temporal transitions.
Although \textbf{Emote} demonstrates visually detailed deformations, its emotion consistency fluctuates across frames, reducing perceived realism.
In contrast, our \textbf{EditEmoTalk} generates both temporally stable and emotionally consistent animations, effectively balancing natural motion, synchronized articulation, and expressive authenticity.
This highlights the advantage of our dual-channel emotion-guided architecture and emotional consistency loss in producing vivid and lifelike 3D talking heads.

\subsection{Ablation Study} To assess the contribution of key components in our framework, we conduct ablation experiments on the MEAD dataset, focusing on (1) the dual-train strategy and (2) the emotional consistency loss $\mathcal{L}_{emo}$. \textbf{Dual-Train Strategy.} When the dual-train strategy is removed, the model cannot fully leverage both original speech-driven embeddings and edited emotion embeddings during training. This results in reduced flexibility for continuous emotion editing and slightly degraded facial motion fidelity, indicating that dual-train effectively balances faithful speech-driven generation with controllable emotion manipulation. \textbf{Emotional Consistency Loss.} Excluding $\mathcal{L}_{emo}$ leads to jittery facial motions and inconsistent emotional intensity across frames. Incorporating this loss enforces temporal coherence and emotion stability, producing smoother and more natural facial dynamics. Table~\ref{tab:ablation} reports quantitative results. Both components contribute substantially to performance improvements, and the full model with dual-train strategy and $\mathcal{L}_{emo}$ achieves the best metrics across Vertex Error (VE), Lip Vertex Error (LVE), Mouth Opening Deviation (MOD), and Facial Dynamics Deviation (FDD).

\section{Conclusion}

We introduce EditEmoTalk, a controllable speech-driven 3D facial animation framework enabling continuous and fine-grained emotion editing. Leveraging a boundary-aware semantic embedding, it models inter-emotion boundaries to construct a continuous expression manifold for smooth, interpretable emotion manipulation. An emotional consistency loss via a lightweight mapping network ensures generated motion faithfully reflects the target emotion. Experiments show that EditEmoTalk achieves superior expressiveness, controllability, and cross-dataset generalization, while maintaining accurate lip synchronization.

%% file: main.bib
@inproceedings{thambiraja2023imitator,
  title={Imitator: Personalized speech-driven 3d facial animation},
  author={Thambiraja, Balamurugan and Habibie, Ikhsanul and Aliakbarian, Sadegh and Cosker, Darren and Theobalt, Christian and Thies, Justus},
  booktitle={Proceedings of the IEEE/CVF international conference on computer vision},
  pages={20621--20631},
  year={2023}
}

@article{jiang2024audio,
  title={Audio-Driven Facial Animation with Deep Learning: A Survey},
  author={Jiang, Diqiong and Chang, Jian and You, Lihua and Bian, Shaojun and Kosk, Robert and Maguire, Greg},
  journal={Information},
  volume={15},
  number={11},
  pages={675},
  year={2024},
  publisher={MDPI}
}

@inproceedings{fan2022faceformer,
  title={Faceformer: Speech-driven 3d facial animation with transformers},
  author={Fan, Yingruo and Lin, Zhaojiang and Saito, Jun and Wang, Wenping and Komura, Taku},
  booktitle={Proceedings of the IEEE/CVF conference on computer vision and pattern recognition},
  pages={18770--18780},
  year={2022}
}

@article{chai2022personalized,
  title={Personalized audio-driven 3d facial animation via style-content disentanglement},
  author={Chai, Yujin and Shao, Tianjia and Weng, Yanlin and Zhou, Kun},
  journal={IEEE Transactions on Visualization and Computer Graphics},
  volume={30},
  number={3},
  pages={1803--1820},
  year={2022},
  publisher={IEEE}
}

@inproceedings{peng2023emotalk,
  title={Emotalk: Speech-driven emotional disentanglement for 3d face animation},
  author={Peng, Ziqiao and Wu, Haoyu and Song, Zhenbo and Xu, Hao and Zhu, Xiangyu and He, Jun and Liu, Hongyan and Fan, Zhaoxin},
  booktitle={Proceedings of the IEEE/CVF international conference on computer vision},
  pages={20687--20697},
  year={2023}
}

@inproceedings{danvevcek2023emotional,
  title={Emotional speech-driven animation with content-emotion disentanglement},
  author={Dan{\v{e}}{\v{c}}ek, Radek and Chhatre, Kiran and Tripathi, Shashank and Wen, Yandong and Black, Michael and Bolkart, Timo},
  booktitle={SIGGRAPH Asia 2023 Conference Papers},
}

@book{thalmann1993models,
  title={Models and techniques in computer animation},
  author={Thalmann, Nadia Magnenat and Thalmann, Daniel},
  year={1993},
  publisher={Springer}
}

@inproceedings{brand1999voice,
  title={Voice puppetry},
  author={Brand, Matthew},
  booktitle={Proceedings of the 26th annual conference on Computer graphics and interactive techniques},
  pages={21--28},
  year={1999}
}

@inproceedings{deena2009speech,
  title={Speech-driven facial animation using a shared Gaussian process latent variable model},
  author={Deena, Salil and Galata, Aphrodite},
  booktitle={International Symposium on Visual Computing},
  pages={89--100},
  year={2009},
  organization={Springer}
}

@article{deena2013visual,
  title={Visual speech synthesis using a variable-order switching shared Gaussian process dynamical model},
  author={Deena, Salil and Hou, Shaobo and Galata, Aphrodite},
  journal={IEEE transactions on multimedia},
  volume={15},
  number={8},
  pages={1755--1768},
  year={2013},
  publisher={IEEE}
}

@article{suwajanakorn2017synthesizing,
  title={Synthesizing obama: learning lip sync from audio},
  author={Suwajanakorn, Supasorn and Seitz, Steven M and Kemelmacher-Shlizerman, Ira},
  journal={ACM Transactions on Graphics (ToG)},
  volume={36},
  number={4},
  pages={1--13},
  year={2017},
  publisher={ACM New York, NY, USA}
}

@article{karras2017audio,
  title={Audio-driven facial animation by joint end-to-end learning of pose and emotion},
  author={Karras, Tero and Aila, Timo and Laine, Samuli and Herva, Antti and Lehtinen, Jaakko},
  journal={ACM Transactions on Graphics (ToG)},
  volume={36},
  number={4},
  pages={1--12},
  year={2017},
  publisher={ACM New York, NY, USA}
}

@article{taylor2017deep,
  title={A deep learning approach for generalized speech animation},
  author={Taylor, Sarah and Kim, Taehwan and Yue, Yisong and Mahler, Moshe and Krahe, James and Rodriguez, Anastasio Garcia and Hodgins, Jessica and Matthews, Iain},
  journal={ACM Transactions On Graphics (TOG)},
  volume={36},
  number={4},
  pages={1--11},
  year={2017},
  publisher={ACM New York, NY, USA}
}

@inproceedings{han2025pestalk,
  title={PESTalk: Speech-Driven 3D Facial Animation with Personalized Emotional Styles},
  author={Han, Tianshun and Zhou, Benjia and Liu, Ajian and Liang, Yanyan and Zhang, Du and Lei, Zhen and Wan, Jun},
  booktitle={Proceedings of the 33rd ACM International Conference on Multimedia},
  pages={7893--7901},
  year={2025}
}

@article{lu2021live,
  title={Live speech portraits: real-time photorealistic talking-head animation},
  author={Lu, Yuanxun and Chai, Jinxiang and Cao, Xun},
  journal={ACM Transactions on Graphics (ToG)},
  volume={40},
  number={6},
  pages={1--17},
  year={2021},
  publisher={ACM New York, NY, USA}
}

@inproceedings{tian2024emo,
  title={Emo: Emote portrait alive generating expressive portrait videos with audio2video diffusion model under weak conditions},
  author={Tian, Linrui and Wang, Qi and Zhang, Bang and Bo, Liefeng},
  booktitle={European Conference on Computer Vision},
  pages={244--260},
  year={2024},
  organization={Springer}
}

@article{xu2024vasa,
  title={Vasa-1: Lifelike audio-driven talking faces generated in real time},
  author={Xu, Sicheng and Chen, Guojun and Guo, Yu-Xiao and Yang, Jiaolong and Li, Chong and Zang, Zhenyu and Zhang, Yizhong and Tong, Xin and Guo, Baining},
  journal={Advances in Neural Information Processing Systems},
  volume={37},
  pages={660--684},
  year={2024}
}

@inproceedings{huang2023collaborative,
  title={Collaborative diffusion for multi-modal face generation and editing},
  author={Huang, Ziqi and Chan, Kelvin CK and Jiang, Yuming and Liu, Ziwei},
  booktitle={Proceedings of the IEEE/CVF conference on computer vision and pattern recognition},
  pages={6080--6090},
  year={2023}
}

@article{melnik2024face,
  title={Face generation and editing with stylegan: A survey},
  author={Melnik, Andrew and Miasayedzenkau, Maksim and Makaravets, Dzianis and Pirshtuk, Dzianis and Akbulut, Eren and Holzmann, Dennis and Renusch, Tarek and Reichert, Gustav and Ritter, Helge},
  journal={IEEE Transactions on pattern analysis and machine intelligence},
  volume={46},
  number={5},
  pages={3557--3576},
  year={2024},
  publisher={IEEE}
}

@inproceedings{patashnik2021styleclip,
  title={Styleclip: Text-driven manipulation of stylegan imagery},
  author={Patashnik, Or and Wu, Zongze and Shechtman, Eli and Cohen-Or, Daniel and Lischinski, Dani},
  booktitle={Proceedings of the IEEE/CVF international conference on computer vision},
  pages={2085--2094},
  year={2021}
}

@inproceedings{jiang2023styleipsb,
  title={Styleipsb: Identity-preserving semantic basis of stylegan for high fidelity face swapping},
  author={Jiang, Diqiong and Song, Dan and Tong, Ruofeng and Tang, Min},
  booktitle={Proceedings of the IEEE/CVF conference on computer vision and pattern recognition},
  pages={352--361},
  year={2023}
}

@inproceedings{blanz1999morphable,
  title={A Morphable Model for the Synthesis of 3D Faces},
  author={Blanz, V and Vetter, T},
  booktitle={26th Annual Conference on Computer Graphics and Interactive Techniques (SIGGRAPH 1999)},
  pages={187--194},
  year={1999},
  organization={ACM Press}
}

@article{li2017learning,
  title={Learning a model of facial shape and expression from 4D scans.},
  author={Li, Tianye and Bolkart, Timo and Black, Michael J and Li, Hao and Romero, Javier},
  journal={ACM Trans. Graph.},
  volume={36},
  number={6},
  pages={194--1},
  year={2017}
}

@inproceedings{jiang2019disentangled,
  title={Disentangled representation learning for 3d face shape},
  author={Jiang, Zi-Hang and Wu, Qianyi and Chen, Keyu and Zhang, Juyong},
  booktitle={Proceedings of the IEEE/CVF conference on computer vision and pattern recognition},
  pages={11957--11966},
  year={2019}
}

@article{hsu2021hubert,
  title={Hubert: Self-supervised speech representation learning by masked prediction of hidden units},
  author={Hsu, Wei-Ning and Bolte, Benjamin and Tsai, Yao-Hung Hubert and Lakhotia, Kushal and Salakhutdinov, Ruslan and Mohamed, Abdelrahman},
  journal={IEEE/ACM transactions on audio, speech, and language processing},
  volume={29},
  pages={3451--3460},
  year={2021},
  publisher={IEEE}
}

@article{ma2023emotion2vec,
  title={emotion2vec: Self-supervised pre-training for speech emotion representation},
  author={Ma, Ziyang and Zheng, Zhisheng and Ye, Jiaxin and Li, Jinchao and Gao, Zhifu and Zhang, Shiliang and Chen, Xie},
  journal={arXiv preprint arXiv:2312.15185},
  year={2023}
}

@inproceedings{kaisiyuan2020mead,
 author = {Wang, Kaisiyuan and Wu, Qianyi and Song, Linsen and Yang, Zhuoqian and Wu, Wayne and Qian, Chen and He, Ran and Qiao, Yu and Loy, Chen Change},
 title = {MEAD: A Large-scale Audio-visual Dataset for Emotional Talking-face Generation},
 booktitle = {ECCV},
 month = Augest,
 year = {2020}
}

@article{cao2014crema,
  title={Crema-d: Crowd-sourced emotional multimodal actors dataset},
  author={Cao, Houwei and Cooper, David G and Keutmann, Michael K and Gur, Ruben C and Nenkova, Ani and Verma, Ragini},
  journal={IEEE transactions on affective computing},
  volume={5},
  number={4},
  pages={377--390},
  year={2014},
  publisher={IEEE}
}

@article{livingstone2018ryerson,
  title={The Ryerson Audio-Visual Database of Emotional Speech and Song (RAVDESS): A dynamic, multimodal set of facial and vocal expressions in North American English},
  author={Livingstone, Steven R and Russo, Frank A},
  journal={PloS one},
  volume={13},
  number={5},
  pages={e0196391},
  year={2018},
  publisher={Public Library of Science San Francisco, CA USA}
}

@inproceedings{zhang2021flow,
  title={Flow-Guided One-Shot Talking Face Generation With a High-Resolution Audio-Visual Dataset},
  author={Zhang, Zhimeng and Li, Lincheng and Ding, Yu and Fan, Changjie},
  booktitle={Proceedings of the IEEE/CVF Conference on Computer Vision and Pattern Recognition},
  pages={3661--3670},
  year={2021}
}

@inproceedings{retsinas20243d,
  title={3d facial expressions through analysis-by-neural-synthesis},
  author={Retsinas, George and Filntisis, Panagiotis P and Danecek, Radek and Abrevaya, Victoria F and Roussos, Anastasios and Bolkart, Timo and Maragos, Petros},
  booktitle={Proceedings of the IEEE/CVF Conference on Computer Vision and Pattern Recognition},
  pages={2490--2501},
  year={2024}
}

@inproceedings{richard2021meshtalk,
  title={Meshtalk: 3d face animation from speech using cross-modality disentanglement},
  author={Richard, Alexander and Zollh{\"o}fer, Michael and Wen, Yandong and De la Torre, Fernando and Sheikh, Yaser},
  booktitle={Proceedings of the IEEE/CVF international conference on computer vision},
  pages={1173--1182},
  year={2021}
}

@inproceedings{xing2023codetalker,
  title={Codetalker: Speech-driven 3d facial animation with discrete motion prior},
  author={Xing, Jinbo and Xia, Menghan and Zhang, Yuechen and Cun, Xiaodong and Wang, Jue and Wong, Tien-Tsin},
  booktitle={Proceedings of the IEEE/CVF Conference on Computer Vision and Pattern Recognition},
  pages={12780--12790},
  year={2023}
}

@inproceedings{cudeiro2019capture,
  title={Capture, learning, and synthesis of 3D speaking styles},
  author={Cudeiro, Daniel and Bolkart, Timo and Laidlaw, Cassidy and Ranjan, Anurag and Black, Michael J},
  booktitle={Proceedings of the IEEE/CVF conference on computer vision and pattern recognition},
  pages={10101--10111},
  year={2019}
}

@article{sun2024diffposetalk,
  title={Diffposetalk: Speech-driven stylistic 3d facial animation and head pose generation via diffusion models},
  author={Sun, Zhiyao and Lv, Tian and Ye, Sheng and Lin, Matthieu and Sheng, Jenny and Wen, Yu-Hui and Yu, Minjing and Liu, Yong-jin},
  journal={ACM Transactions on Graphics (TOG)},
  volume={43},
  number={4},
  pages={1--9},
  year={2024},
  publisher={ACM New York, NY, USA}
}

@inproceedings{kim2025deeptalk,
  title={DEEPTalk: Dynamic Emotion Embedding for Probabilistic Speech-Driven 3D Face Animation},
  author={Kim, Jisoo and Cho, Jungbin and Park, Joonho and Hwang, Soonmin and Kim, Da Eun and Kim, Geon and Yu, Youngjae},
  booktitle={Proceedings of the AAAI Conference on Artificial Intelligence},
  volume={39},
  number={4},
  pages={4275--4283},
  year={2025}
}

@inproceedings{peebles2023scalable,
  title={Scalable diffusion models with transformers},
  author={Peebles, William and Xie, Saining},
  booktitle={Proceedings of the IEEE/CVF international conference on computer vision},
  pages={4195--4205},
  year={2023}
}

@article{feng2025faceedittalker,
  title={FaceEditTalker: Interactive Talking Head Generation with Facial Attribute Editing},
  author={Feng, Guanwen and Ma, Zhiyuan and Li, Yunan and Jing, Junwei and Yang, Jiahao and Miao, Qiguang},
  journal={arXiv preprint arXiv:2505.22141},
  year={2025}
}

@inproceedings{liang2024cstalk,
  title={Cstalk: Correlation supervised speech-driven 3d emotional facial animation generation},
  author={Liang, Xiangyu and Zhuang, Wenlin and Wang, Tianyong and Geng, Guangxing and Geng, Guangyue and Xia, Haifeng and Xia, Siyu},
  booktitle={2024 IEEE 18th International Conference on Automatic Face and Gesture Recognition (FG)},
  pages={1--5},
  year={2024},
  organization={IEEE}
}

@article{jafari2024jambatalk,
  title={JambaTalk: Speech-Driven 3D Talking Head Generation Based on Hybrid Transformer-Mamba Model},
  author={Jafari, Farzaneh and Berretti, Stefano and Basu, Anup},
  journal={arXiv preprint arXiv:2408.01627},
  year={2024}
}

@article{ji2024realtalk,
  title={Realtalk: Real-time and realistic audio-driven face generation with 3d facial prior-guided identity alignment network},
  author={Ji, Xiaozhong and Lin, Chuming and Ding, Zhonggan and Tai, Ying and Zhu, Junwei and Hu, Xiaobin and Luo, Donghao and Ge, Yanhao and Wang, Chengjie},
  journal={arXiv preprint arXiv:2406.18284},
  year={2024}
}

@inproceedings{stan2023facediffuser,
  title={Facediffuser: Speech-driven 3d facial animation synthesis using diffusion},
  author={Stan, Stefan and Haque, Kazi Injamamul and Yumak, Zerrin},
  booktitle={Proceedings of the 16th ACM SIGGRAPH Conference on Motion, Interaction and Games},
  pages={1--11},
  year={2023}
}
